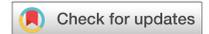

OPEN

# Polarised crowd in motion: insights into statistical and dynamical behavior

Pratikshya Jena✉ & Shradha Mishra

The collection of active agents often exhibits intriguing statistical and dynamical properties, particularly when considering human crowds. In this study, we have developed a computational model to simulate the recent experiment on real marathon races by Bain et al. (Science 363:46–49, 2019). Our primary goal is to investigate the impact of race staff on crowd dynamics. By comparing simulated races with and without the presence of race staff, our study reveals that the local velocity and density of participants display a wave pattern akin to real races for both the cases. The observed traveling wave in the crowd consistently propagates at a constant speed, regardless of the system size under consideration. The participants' dynamics in the longitudinal direction primarily contribute to velocity fluctuations, while fluctuations in the transverse direction are suppressed. In the absence of race staff, density and velocity fluctuations weaken without significantly affecting other statistical and dynamic characteristics of the crowd. Through this research, we aim to deepen our understanding of crowd motion, providing insights that can inform the development of effective crowd management strategies and contribute to the successful control of such events.

Comprehending the behaviour or dynamics of the human crowd is valuable for optimizing both our everyday professional operations and our individual well-being. This understanding can significantly enhance managerial effectiveness as well. Events like sports competitions, exhibitions, and national celebrations exemplify the growing demands for appropriate technological solutions and support to effectively monitor and manage large crowds[1–4]. To prevent stampedes[5,6] and ensure safe and enjoyable events, crowd monitoring has become essential. Numerous studies are available on crowd management that include real-world events involving camera tracking and data collection, as well as computer-simulated scenarios to investigate crowd behaviour[5,7–28].

Another prominent event is the marathon race, which demands special attention when addressing crowd control measures. Indeed, there are numerous studies focused on city marathons[29–35], which aim to capture and analyse crowd dynamics during the races where these studies utilize various methods, including video analysis, data tracking, and participant surveys to understand how the crowd behave and interact during marathon events. The computer-simulated numerical models[36–39] for marathon attempt to replicate the real events. These models use mathematical equations and algorithms to represent the interactions and dynamics of the crowd during the marathon.

In a recent study[34], the authors have utilized tens of thousands road-race participants in four starting corals: Chicago 2016, Chicago 2017, Paris 2017, Atlanta 2017[34] to explain the flowing behaviour of polarized crowds by examining its response to boundary motion. The outcomes of this experimental observations as well the continuum description[34], elucidates the crowd dynamics specifically concerning velocity and density waves and show that the stimulations from side boundaries are inefficient and that optimal information transfer is obtained from guiding the crowd from its forefront. The longitudinal velocity wave propagates upstream at a constant speed which is a characteristics of information transfer in the polarized crowd in all races. However, the orientational fluctuations are suppressed locally in transverse direction.

Through this work, our objective is to develop a microscopic agent-based model that can be applicable to polarized crowds. Here, we consider a collection of participants of the crowd on a two-dimensional track. The position and velocity of the individual participant are updated using agent-based equations of motion to capture the essential dynamics of runners during a marathon event. Motivated by the real-world situations where the crowd can be guided by some forefront (race staff)[34] or completely independent crowd free to move on a desired track, we provide the model with and without race staff at the forefront. In large races, the race staff form boundary to monitor the crowd motion and spectator areas, ensuring safety and the prevention of interference among the participants[40]. The races without race staff may lack proper organization and management but are still enjoyable with a higher degree of freedom[22,25].

Department of Physics, IIT(BHU), Varanasi 221005, U.P., India. ✉email: pratikshyajena.rs.phy20@itbhu.ac.in

The principal results of this paper are the propagation of hybrid coupled velocity-density wave throughout the system opposite to the crowd motion, as observed in recent experimental work[34]. The appearance of such coupled velocity-density wave is not limited to the participants who are macroscopic in size, but also present on the mesoscopic scale; like the collective motion of bacteria. In the work of[41], it is found that the environmental topology of complex structures is used by Escherichia coli to create traveling waves of high cell density, a prelude to quorum sensing. Additionally, the speed distribution depicts the propagating wave has a constant speed irrespective of the number of participants in the races considered. Based on the observation from the distribution for longitudinal and transverse velocity, it is found that the velocity shows the large fluctuations in longitudinal direction, whereas the fluctuations are suppressed in transverse direction. The density and velocity propagate periodically through the system.

## Model and numerical methodology

We model a collection of participants on a two-dimensional passage. We use the agent based dynamical equations of motion to describe the motion of runners. It takes into account the position $r_i = (x_i, y_i)$ and velocity $V_i = (V_{xi}, V_{yi})$ of the $ith$ participant. We choose the $x-$ and $y-$ directions as direction parallel $\parallel$ (longitudinal) and $\perp$ (transverse) to the direction of moving crowd respectively as shown in the schematic of the model Fig. 1. The velocity of each participant is updated by;

$$\frac{dV_i}{dt} = \boldsymbol{h} + \mu_1 F_i + \bar{\eta}_i\,(\boldsymbol{r}, \boldsymbol{t}) \tag{1}$$

where left hand side is simply the inertial term, the mass of the participant is taken as unity (because the Gravity is unimportant for the motion in a plane). The three terms on the right-hand side are different forces: (i) biased drive to polarise the crowd along the track of the path $+x$direction with $\boldsymbol{h} = (h_0, 0)$, (ii) The short ranged soft-repulsive interaction among the participant denoted by $F_i = \sum_{j=1}^{N} f_{ij}$, where $f_{ij} = k\,(r_{ij} - 2r_0)\,\widehat{r_{ij}}$, if $r_{ij} \leq 2r_0$ otherwise it is zero. Such force accounts for the mutual exclusion among the participants of size $r_0$, here $r_{ij} = |r_j - r_i|$, $\widehat{r_{ij}} = r_j - r_i$. The mobility $\mu_1$ and strength of the force $k$ are chosen such that the elastic time scale $(\mu_1 k)^{-1/2}$ is set to one and (iii) small random uncorrelated thermal noise $\bar{\eta}$ having strengths $(\Delta \eta_x, \Delta \eta_y)$ taking care of perturbation arises due to any kind of random fluctuations present in the system. The motivation behind the three terms on the right-hand side of Eq. 1 is mainly with the observation of the crowd motion in the animation of[34] and the force term in the hydrodynamic description of[34]. The only relevant force on the moving crowd as given in[34] is the acceleration in the longitudinal direction that makes the crowd polarised. Further, particles move in direction of their velocity obeying the following rule;

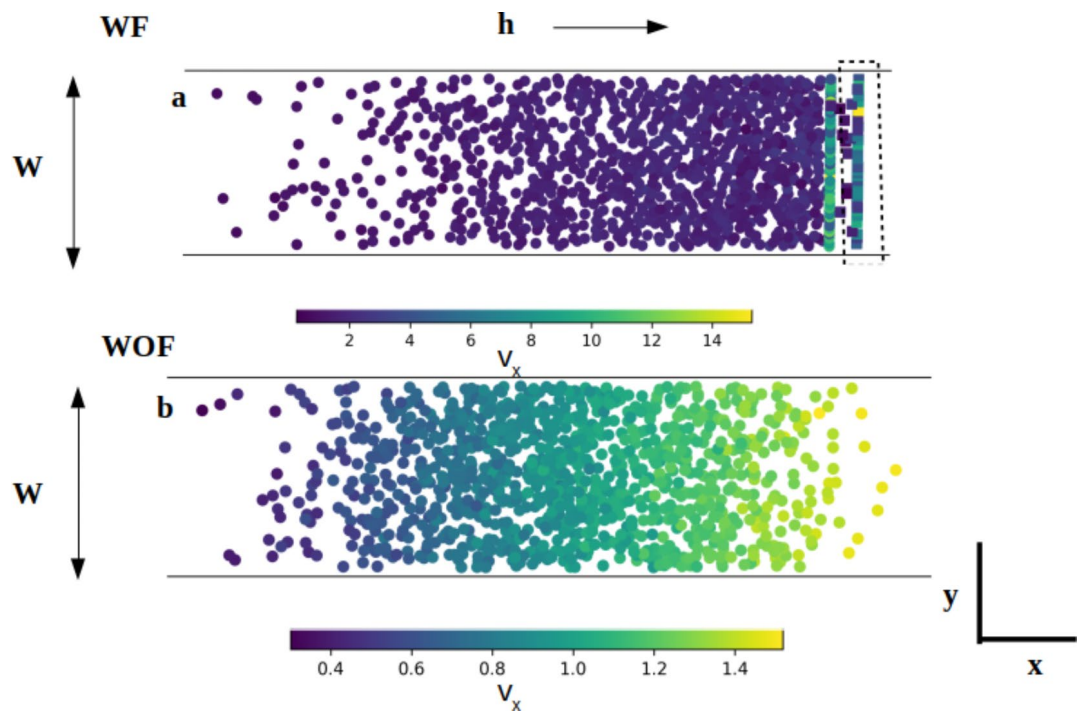

**Fig. 1.** The schematic of our model for the races (**a**) with frontliners and (**b**) without frontliners. The images are generated from the simulation. The circles depict the participants in the races in both the scenarios and colour of the circle represent the local $x-$ component of velocity $V_x$ of the participants. **h** is the external drive to polarize the crowd. In Figure (**a**), the squares inside the rectangle represent the frontliners. $W$ is the width of the track in the transverse direction.

$$\frac{dx_i}{dt} = v_0 P_{xi} + \zeta_{xi}(t) \quad (2)$$

$$\frac{dy_i}{dt} = P_{yi} + \zeta_{yi}(t) \quad (3)$$

In the $y-$direction participants simply follow the velocity in $y-$ direction, whereas in the $x-$direction the step size of the participants $v_0$ is modified using a Gaussian distribution $P(v_0) = \frac{1}{\sigma\sqrt{2\pi}}\exp\left(\frac{v_0-\mu}{2\sigma}\right)^2$ with mean $\mu=0.6$ and variance $\sigma=0.03$, where at every time and for different participants it is obtained independently. Motivated with the recent experiments on the dense bacterial solution[42,43], we introduced the distribution of speed $P(v_0)$ of the participants in the longitudinal direction. As reported in[43], the range of microswimmers maintained a common distribution in their speed, moreover in[42], it was found that the collective migration increases in bacterial solution by moving slowly. Thus, introducing a distribution in speed, where we have finite number of participants moving slowly may help the crowd propagation. Also introducing a distribution of speed in longitudinal direction avoids overcrowding. In the transverse direction, since the mean of the velocity is always zero, no such modification is required. The similar mechanism can also be obtained by introducing a distribution for $h_0$. In both directions, velocity is further modified with an additional random uncorrelated noise $\bar{\zeta}$ of strengths $(\Delta\zeta_x, \Delta\zeta_y)$ to account for any kind of random fluctuations present in the system. For comparison we also checked the model with constant speed with $v_0$. Most of the detailed results of the manuscript is for the speed distribution $P(v_0)$.

In the experiment[34] the participants in the race are guided towards the starting line by chains of staff members or frontliners. To incorporate the presence of such race staff, we introduce race staff in our model, which act like moving forefront for the crowd. To introduce the frontliners, we marked 50 participants present at the forefront of the crowd as frontliners when race starts, shown inside the rectangular box in Fig. 1a. The velocity and position update of the participants and the frontliners are as given in Eqs. (1, 2 and 3). The frontliners have no constraints in their dynamic in the longitudinal direction, whereas the participants experience the frontliners as moving wall at their front and cannot cross it. We also studied the race without frontliners; by treating all the participants equivalent and they do not feel any moving forefront as shown in Fig. 1b. To define the front of the crowd in this case we choose first 50 participants, which are at the front of the crowd (largest $x-$coordinates) and take the mean of their $x-$ coordinates. The same mean is defined as $x_0$ or front of the crowd. The two systems in the presence and absence of frontliners are named as system with and without frontliners (WF and WOF respectively).

We choose effective size of the participants $r_0 = 0.8$ as the intrinsic length scale in the system and the ratio of $r_0/\mu = 0.8/0.6 = 4/3 = \tau$ is the intrinsic time scale in the system. This is a typical time on an average a participant takes to cross its own size. Further we re-scaled all the lengths and times in the system in terms of $r_0$ and $\tau$. We define a dimensionless acceleration $\bar{h} = h_0\tau^2/r_0$. We consider races with varying numbers of participants ($N$) within the crowd. Specifically, we choose values of $N$ in the range 500–7000 (typical number of participants in a standard race). These different numbers allow us to examine the impact of crowd size on the observed dynamics. For all $N$ values, the participants are initially placed in narrow width $W = 62r_0$ in the transverse direction with reflecting boundaries. Initial number density $\rho = N/(W \times L_0)$ of the participants is 1.0, where $L_0$ is the initial spread of the participants in the longitudinal direction and it is varied from 12.5 $r$ to 125 $r_0$ for different size of the races. Later, the participants are allowed to move according to the update equations given in Eqs. (1, 2 and 3). We have performed the simulation by taking the small-time step $\Delta t = \frac{2}{3} \times 10^{-3} \tau$. One simulation step is counted when all the participants as well as the frontliners are updated once. The simulation is performed for total time $T = 500\tau$. Motion in longitudinal direction is in open space, and there is no direct attraction among the participants, hence spread of the crowd in the longitudinal direction $L(t)$ increases with time. The spread is faster for the system WOF in comparison to system WF. Therefore, our most of the measurements are performed for time up to $t = 300\tau$, such that density is not very low and interaction among the participants is relevant. In the experiment[34] as well the crowd motion is observed for 100 to 500 s such that a finite mean density is maintained in the system. The stationary state of the system is defined when the distance between the participants is large and they become non-interacting. Most of our observations are for time much smaller than the time it takes to reach the non-interacting limit.

We choose the strengths of the four random fluctuations $\Delta\eta_x = \Delta\eta_y = \Delta\zeta_x = \Delta\zeta_y = 10^{-4}$ small. The dimensionless acceleration $\bar{h}$ is fixed to 2/9, by fixing $h_0 = 0.1$. Hence, participants experience a constant flow in the direction of **h**. Averaging over 50 to 100 independent realisations are performed for good statistics.

## Results

We first analyse the impact of frontliners on the dynamics of participants in the races. Movies [SM1] and [SM2] capture the animation of the system WF and WOF for a particular crowd size by setting $N = 1000$ respectively. The corresponding figures at time $t \simeq 100\tau$ are shown in Fig. 1a,b. The circles represent the participants and the colours of the circles represent the magnitude of x − component of velocity $V_x(t)$ of the participants. We can very clearly observe that starting from the initial homogeneous and random velocity of particles, density and velocity patterns are formed. The detail results of the two systems with and without frontliners will be discussed later. The similar animation for the system WF for constant $v_0$ is provided in Movie [SM3]. The basic appearance of density and velocity waves can be seen in this case too.

Based on the experiments on the real races[29,34,37,44], various observable such as velocity and density waves, speed and velocity distribution of the participants, distribution of density in the system as well as density and velocity auto-correlation can be examined to explore the dynamics and characteristics of the moving crowd. These observables are valuable tools for characterizing the dynamic behaviour of participants during the races.

To get the results from our model, we assume the crowd as a continuum and investigate the dynamics of the crowd by measuring the local coarse-grained density $\rho(x,t)$ and velocity $v_x(x,t)$. The movement of individual participants during the races generates pressure, leading to emergence of hybrid wave of velocity-density which propagates through the system. We further examine the participants speed ($u = \sqrt{(V_x^2 + V_y^2)}$) distribution P(u) and velocity distribution $P(V_{i,j})$ (where $(i,j)$ represent x and y components of velocity) to understand the characteristics of the travelling wave. To interpret the spreading and squeezing of crowd during the races, we also analyze the density distribution P(ρ) of the crowd in races having different numbers of participants and at different times. This can be beneficial to understand the crowd distribution to manage and control the stampede like situation during the marathon events. We additionally calculate the velocity and density auto-correlation functions $C_{v_x}(t)$ and $C_\rho(t)$ respectively to get the information about how the velocity and density are correlated with respect to time. The decay of the velocity auto-correlation provides information about characteristic time-scale associated with the relaxation of the system. Below the discussion of each observable are provided one by one from the result of our model.

## Velocity and density profiles

To employ a continuum approach to analyse the large-scale motion within the crowd, we plot the local coarse-grained density, denoted as, and the velocity at different spatial positions and time. To define the coarse-grained local density and velocity, we divide the whole track on which the race goes on into $n=500$ rectangular cells of size $\Delta x=10$ in x−direction and breadth equal to the full length of the track in y−direction. The x=0 line is always fixed at the mean of the x−coordinates of the frontliners ($x_0$). Later, the distances in the longitudinal direction for both the model's system WF and WOF are measured with respect to mean x−coordinates of the frontliners and front of the wave respectively. Then we calculate the density in each cell by counting the number of participants in each cell and dividing by the area of the cell. To determine the coarse-grained velocity, we add the longitudinal velocities of the participants in the cell. Figure 2 shows the schematic of the above procedure explained to determine the coarse grain density and velocity.

Figure 3 provides a visual representation of crowd flow by showing the propagation of coarse-grained density $\rho(x,t)$ and velocity $v_x(x,t)$ at three times. The direction of flow of crowd is downward and vertical spread of the track is in the horizontal direction. In Fig. 3a–c for the local density $\rho(x,t)$ and (d-f) for the local velocity $v_x(x,t)$ at three different times $t \simeq (30,75,150)$ τ respectively for the system with frontliners. Similarly, Fig. 3g–i for the local density $\rho(x,t)$ and (j-l) for the local velocity $v_x(x,t)$ at three different times $t \simeq (30,75,150)$ τ respectively for the system without frontliners. Starting from the early time with a constant density in a small region, the density and velocity profile undergoes a propagation inferring a hybrid coupled wave of density and velocity transmitting within the system. It is clearly noticeable that with time an initial homogeneous density of participants splits into density wave, which moves in the direction opposite to the direction of crowd. Here, we give in brief the mechanism of formation of density wave in the system. For the system WF, near the frontliners, the participants cannot cross the frontliners, hence although they have a force in the longitudinal direction, the contribution of force will be small. The dynamic of the participants is only due to the repulsive force and random noises ζ′ s. Additionally, there is a high-density region in the backward direction, that limits the motion in the backward direction and the motion seems to happen primarily in the transverse direction. That controls the crowding at the front of the race. But, since the longitudinal acceleration is active, the participants have to push the crowd in backward direction to get the space to move in the forward direction. That will result in the formation of density wave in the system for the system WF. For the system WOF, the mechanism is the same, the only difference is that the participants have no constraints in the front direction, hence the density wave will be weaker. The same is observed in our simulation as can be seen from the two animations [SM1] and [SM2]. Here, we want to emphasise that the presence of initial density and velocity wave as shown in Fig. 3 is not solely due to the initial condition, but also system's confinement and constrained dynamics plays an important role too. Thus, the characteristics of the system is different from usual chaotic systems, where the non-linear nature of the system is responsible for the initial condition dependent steady state[45].

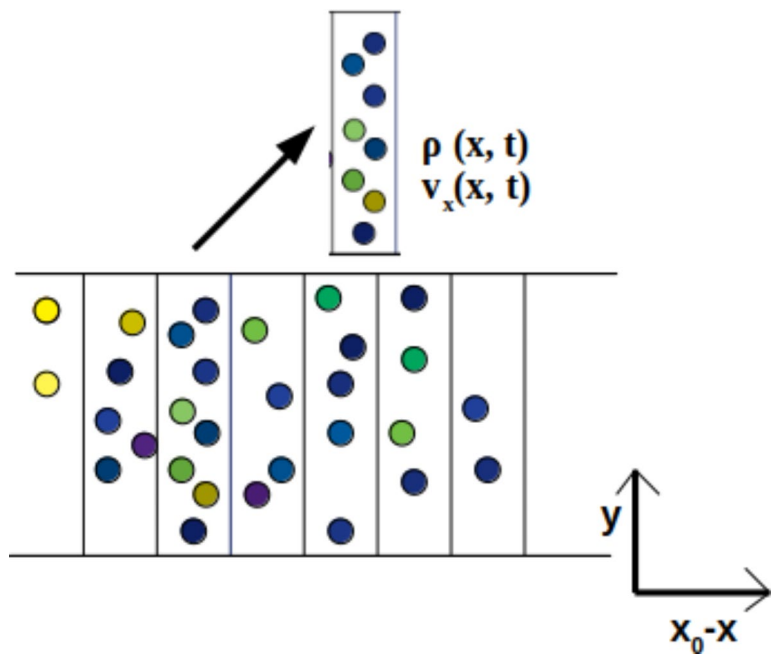

**Fig. 2.** The picture depicts the track is divided into n rectangular cells. Here, the x-axis represents $x_0 - x$, where $x$ and $x_0$ are the coordinates of participants and the mean x-coordinates of the frontliners respectively. The zoomed picture presents a single cell inside which the coarse-grained density $\rho(x,t)$ and velocity $v_x(x,t)$ are calculated. The colour of the circles has the same meaning as in Fig. 1.

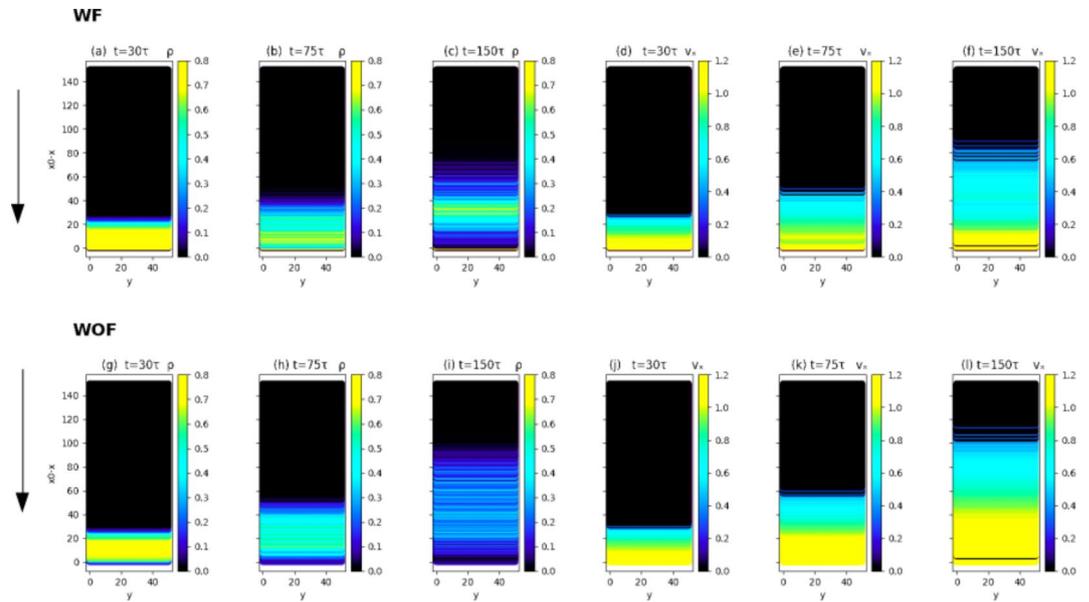

**Fig. 3**. The plot (**a**–**c**) and (**g**–**i**) showcase the coarse-grained density $\rho\,(x,t)$ at three times t = (30,75,150) τ in the presence and absence of frontliners respectively. The plot (**d**–**f**) and (**j**–**l**) represent coarse-grained velocity $v_x\,(x,t)$ for the same. The vertical downward arrow represents the direction of crowd propagation.

In Fig. 4A,B, we plot the one-dimensional coarse-grained density $\rho\,(x,t)$ and velocity $v_x\,(x,t)$ for four different times t ≃ (15,30,45,60) τ for the system WF and WOF respectively. Distinctly, the density pattern shows a peak and the peak propagates in the direction opposite to the direction of crowd motion (as shown by arrow in the figure). With the vertical dashed line, we mark the position of the peak of density and similarly for the velocity (excluding the large spike near the frontliners) for the system WF. For the system WOF, velocity does not show a peak, but a vertical dashed line is drawn at the position where velocity shows an abrupt decay to zero. We find that both the density and the velocity waves travel in the same direction with some phase difference [also see the corresponding animations for the particles and coarse-grained density and velocity in [SM1] and [SM2] for the system WF and WOF respectively]. We find that the x − component of velocity $v_x\,(x,t)$, shows the abrupt decay near the frontliners as shown in Fig. 4A(e–h). A dip in the x − component of velocity near the frontliners due to the suppressed longitudinal component of velocity because of the presence of constraints at the front of the race and high density at the back. Further for the system WOF as shown in Fig. 4B(e–h), we find almost linear decay of $v_x\,(x,t)$, that is due to the weaker density wave in this case and particles at the front of the wave, are moving with relatively larger speed compared to the backward particles, due to the presence of high density around them. To get the more depth of the density and velocity field in the system in Fig. 11(a–d) in the Appendix A we show the plot of density and velocity spatial correlations in the system with and without frontliners at different times.

To further find the relation between density and velocity, in Fig. 5a,b we showcase the local density $\rho\,(x,t)$ vs. local velocity $v_x\,(x,t)$ at three different times t ≃ (15,75,180) τ for the race having 1000 participants in the presence and absence of frontliners respectively. The relation between density and velocity field is highly non-monotonic in nature. For small velocity: first, density increases with $v_x\,(x,t)$ indicating the tail of the crowd which is away from the forefront. This part of the crowd is also shown by the region left to the blue shaded region in the one-dimensional plot of density and shown in Fig. 4A,B. On further increasing velocity, relatively large number of participants move with a moderate velocity, where the plot shows a peak, that mimics the density band in the system. The same is shown by small shaded region with blue in Fig. 4A,B. On further increasing velocity, density again suppresses, but the plot shows the abrupt change for system with frontliners 5(a), whereas the change is gradual for system WOF. For the system WF, there is a thin layer of participants move with high velocity are responsible for such behaviour. The density-velocity plot is narrow for the system WF, whereas it is wider for the system WOF, which can also be seen by the one-dimensional plot of density in Fig. 4A,B, which is narrow for system WF and wider for the system WOF. As we increase time, the nature of density vs. velocity plot remains the same, but slowly crowd disperse and both density and velocity shifts towards the smaller values. This non-monotonic nature is intrinsic to the collective behaviour of participants moving in different races as found in previous experiments[29]. The presence of band of high density is responsible for such non-monotonic behaviour. The plots shown in Fig. 5 gives the local correlation of density and velocity in the system at different times. We also characterised the global correlation of both the field as a function of time by calculating the cross correlation of function $C_{v_x \rho}(t)$ shown in Fig. 12a,b in the Appendix A.

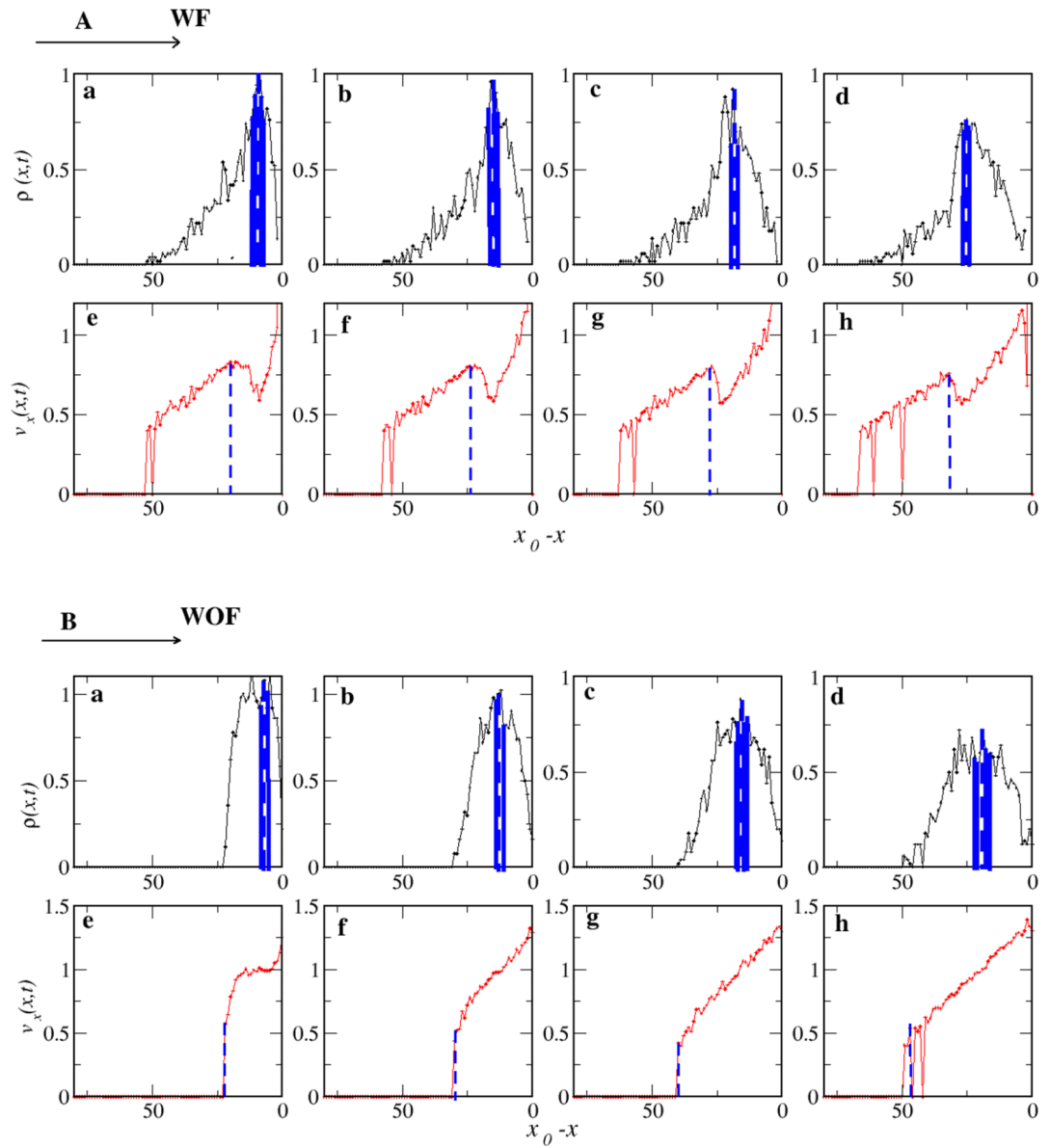

**Fig. 4**. (**A**) The plots (a–d) illustrate the density propagation at different times t = (15,30,45,60) τ, while the plots (e–h) depict the corresponding propagation of velocity at those specific moments, for the systems with frontliners (WF). (**B**) The plots (a–d) illustrate the density propagation, while the plots (e–h) depict the propagation of velocity for the systems without frontliners (WOF) at the exact instances for the with frontliners (WF) scenario. The blue shaded region in all plot shows the region of high-density band, with location of maximum density is represented by vertical dashed line in Fig. **A**(a–d) and **B**(a–d). The vertical dashed line in Fig. **A**(e–h) shows the location of secondary maxima of velocity $v_x(x,t)$ for system WF. In Fig. **B**(e–h), the vertical dashed line depicts the location where velocity shows abrupt change for the system WOF. The arrow represents the direction of crowd propagation in both the cases.

Till now, we focused on the dynamic nature of the crowd, density and velocity pattern in the system. Now, we focus our attention on the characteristics of the moving participants.

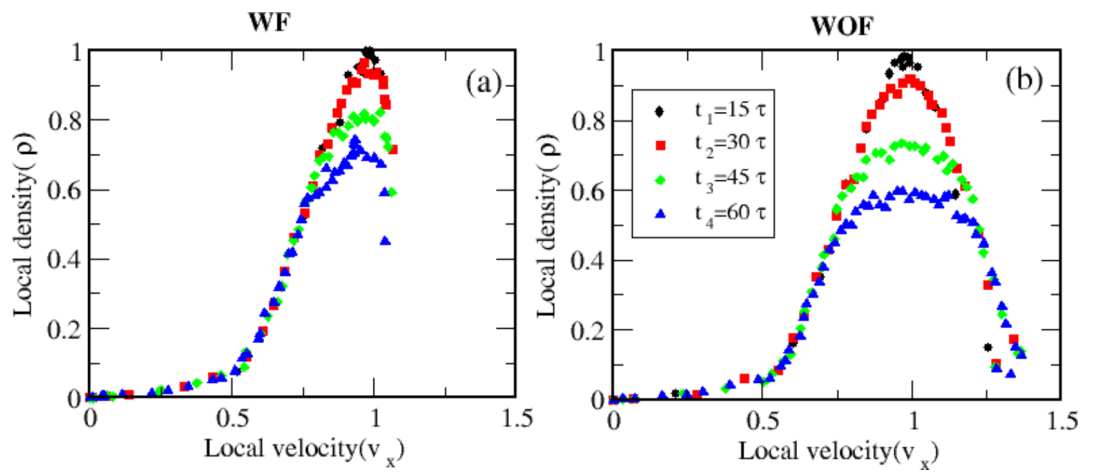

**Fig. 5.** The plots (**a**, **b**) showcase the local density ($\rho$) vs. local velocity ($v_x$) at time t = (15,30,45,60) τ for the race having 1000 participants in the presence and absence of frontliners respectively.

### Speed distribution

The Fig. 6 illustrates the distribution of speeds u of participants for both the cases, the presence and the absence of frontliners. The distribution of local speed P(u) is obtained by calculating distribution of speeds of all the participants. Figure 6a depicts the speed distribution of participants in the presence of frontliners at a specific time t ≃ 225τ. The figure showcases the impact of different system sizes ($N$ = 1000, 1500, 2500, 3000) represented by distinct colours. In the inset-1 of Fig. 6a, we show the lines are the fit to the log-normal function, (where a = 0.4, b = 0.65 for $N$ = 1500). The mean of the speed distribution is non-zero for all N's. The crowd motion is happening through the system with almost same mean speed ranges from (0.5–0.8) for different system sizes. In the inset-2 of Fig. 6a and inset of Fig. 6b we show the plot of mean speed with respect to system size N and find no clear dependence on the system size. We also checked the speed distributions for other times t ≃ 100τ and t ≃ 150τ as shown in Fig. 6c,d for the system with and without frontliners respectively for $N$ = 1000. It is observed that the nature of the distribution as well as the mean remains the same with respect to time. Therefore, wave remains non-dispersive with respect to time in the system unlike the normal waves in any media[46]. The non-dispersive nature of wave we reported here can be compared with the non-dispersive wave in[34], where the nature of the wave remains invariant with respect to time and size of the race. Hence, initially moving crowd relaxes quickly and then it propagates with a constant speed.

In Fig. 6b, the distribution P(u) is shown for the scenario without frontliners. The distribution is wider in comparison to the system WF, with no tail at larger speed. Once again as found for the system WF, the mean speed for all sizes of the race lies in the small range (0.7–0.9). This suggests that the presence or absence of frontliners does not have a significant impact on the mean of the travelling speed of the wave indicating system's robustness in response to external influencers/perturbations. The two distributions of the system WF and WOF have one key difference that the P(u) for the system WF is log−normal and have a long tail, whereas no such tail is observed for system WOF. For the system with frontliners few participants are moving with relatively higher speed. Participants adjacent to the frontliners are mainly contributing for such high speeds [see [SM1] for a visual demonstration of the occurrence].

### Velocity distribution

Further, the fluctuations in the velocity of participants in longitudinal and transverse directions are measured by calculating the distribution of the two components of velocities of participants $V_x$ and $V_y$ for three different sizes of the crowd. Figure 7a,b depicts the longitudinal and transverse component of crowd's velocity distribution for a specific time t = 225τ for the system with frontliners and for the three different system sizes ($N$ = 1500,3000). Similarly, (c-d) represents the identical plots without fronliners in the same sequence. From Fig. 7, it can be observed that the transverse component has zero mean, whereas the longitudinal component exhibits a peak at some non-zero value. These findings suggest that there is no such significant movement in the transverse direction and the propagation of velocity wave primarily occurs in the longitudinal direction. The distribution is narrow for the transverse direction, whereas it is broader for the longitudinal direction suggesting that the moving crowd has larger fluctuations in longitudinal direction in comparison to transverse direction, which is completely opposite to what has been observed for the polar flocks[47–51]. The large fluctuations in longitudinal direction are also observed in real marathon races[34], whereas we expect the suppression of velocity fluctuations in the longitudinal direction, due to the presence of a constant driving force **h**. The confinement present in the transverse direction might be responsible for the suppressed fluctuation. The result of increase in the size of the flock increases the role of confinement and lead to narrower distribution of velocity in the transverse direction as can be clearly seen in Fig. 7a–d.

The two distributions $P(V_x)$ and $P(V_y)$ have some overlap for the system with frontliners, whereas there is no overlap for system without frontliners. This is due to the presence of strong density band, which moves backward and slow down the motion of the participants near to it for the system WF. This can be also seen from

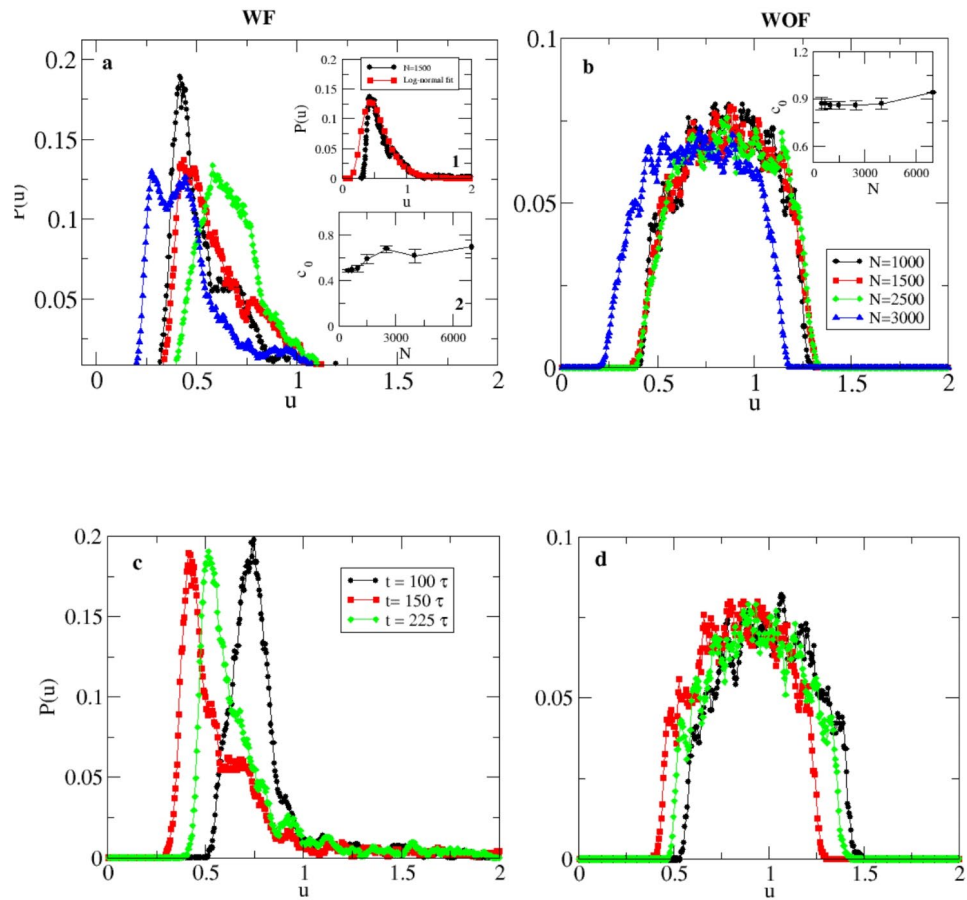

**Fig. 6.** The plots (**a**, **b**) depict the speed distribution of the participants at time t = 225τ for four different system sizes $N = 1000, 1500, 2500, 3000$. The inset-1 and inset-2 of plot (**a**) shows that the speed distribution of participants with frontliners fits well with log-normal distribution and the plot for mean speed $c_0$ vs. N for the system WF. The inset of plot (**b**) shows mean speed $c_0$ vs. N for the system WOF. (**c**, **d**) shows the plot of the speed distribution for fixed system size $N = 1000$ and for three different times t = (100, 150, 225) τ for the system WF and WOF respectively.

Fig. 4A(a–d): density shows a long-tail in the direction opposite to the direction of crowd propagation, and participants moving there have smaller speed as shown in Fig. 4A(e–h). But, for the system WOF, both density and velocity drop abruptly. Since, the intensity of the density band is dilute for system WOF, hence we do not see small velocity of participants in this case and no overlap. The distribution $P(V_x)$ also shows a long tail, that is mainly due to the presence of a moving forefront in the system with frontliners and it results in bigger range of $V_x$ in the system. Restricted motion of the participants due to the presence of frontliners in the direction of moving crowd leads to relatively less broadness for the distribution in longitudinal direction in comparison to the participants without frontliners. That further lead to an overlap in the two distributions for the system with frontliners.

### Density distribution

To further quantify the presence of band of high density and its stability over different realisations, we plot the distribution of local density P(ρ) in the system. Figure 8a–c and d–f, illustrates the density histogram of the system at three different times t ≃ (15, 45, 60) τ, comparing the two cases, the presence and absence of frontliners. As shown in Fig. 8a, at the early time the initial distribution is wide with a large range of density with a bimodal character, represents the presence of a very narrow band with density almost close to the density of frontliners (as shown by the vertical line in each panel). The distribution contracts with time in Fig. 8b,c and retains its bimodal nature as well as the difference from the density close to frontliners density increases. Figure 8d–f depicts the same for time steps t ≃ (15, 45, 60) τ for the system without frontliners. For this case, we see more clear bimodal distribution of P(ρ) for all times. Which is due to the relatively wider band in the system WOF in comparison to the WF. For both the cases, the distribution shrinks to the lower density with respect to time, due to the spread of the crowd in the longitudinal direction.

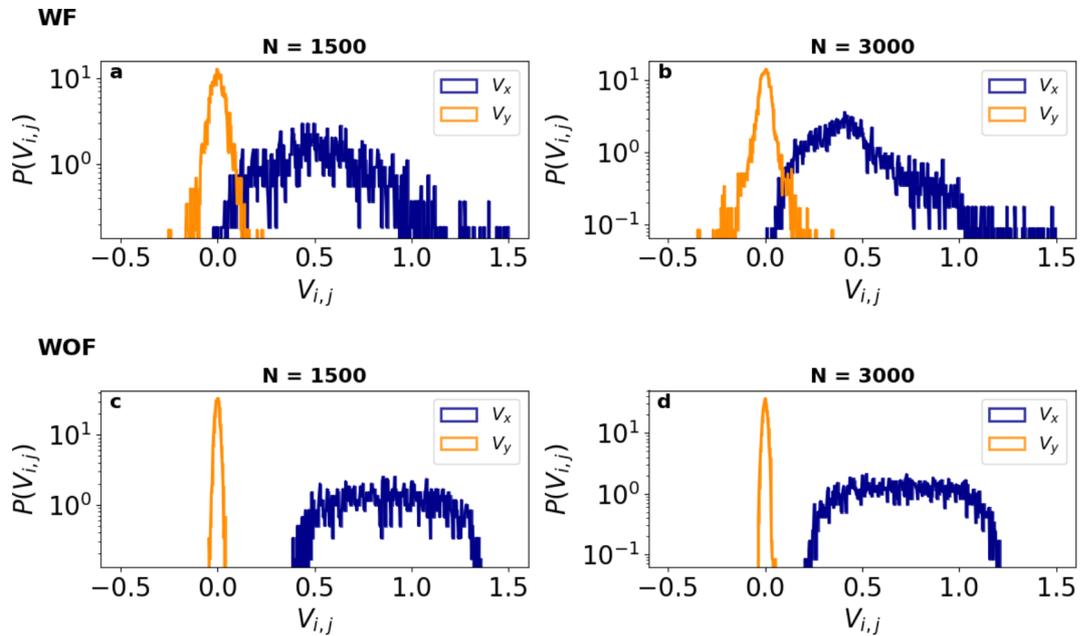

**Fig. 7.** The plot (**a**, **b**) illustrates distribution of longitudinal and transverse components of crowd's velocity P ($V_{i,j}$) at a particular instant (t $\simeq$ 225$\tau$) for different $N$ = 1500 and 3000 respectively in the presence of frontliners. The plot (**c**, **d**) shows replicated plots in the absence of frontliners.

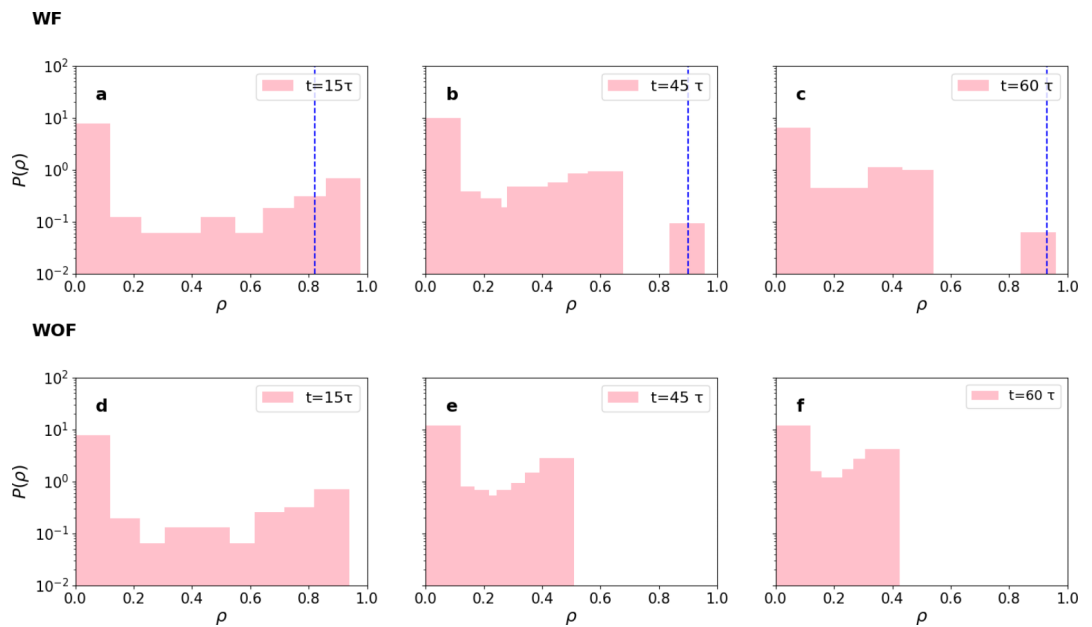

**Fig. 8.** In the plots (**a**–**c**), the histograms display the density at three different times t = (15,45,60) $\tau$ for a race with $N$ = 3000 individuals in presence of the frontliners. Here the x−axis represents the local $\rho$ and the y−axis represents the distribution of density P($\rho$). The blue dotted lines represent the density of the frontliners. The plots (**d**–**f**) showcase the same in the absence of frontliners.

Further, we also calculated the P($\rho$) at fixed time and for different system size $N$ = 1000, 1500, 2500 for both the cases system WF and WOF as shown in Fig. 9a–c and d–f respectively. For the system WF, the P($\rho$) shows the bimodal structure for all system sizes. Also, for the system WOF, the nature of the distribution remains the same for all system size, it shows two small peaks at low and high density with a dip for intermediate density. The bimodal nature becomes more prominent for system WOF, due to wider band in comparison to narrow band for the system WF. This implies the characteristics of the density remains invariant with respect to the system size. The one small bar at high density for the system WF shows the density of participants close to the frontliners.

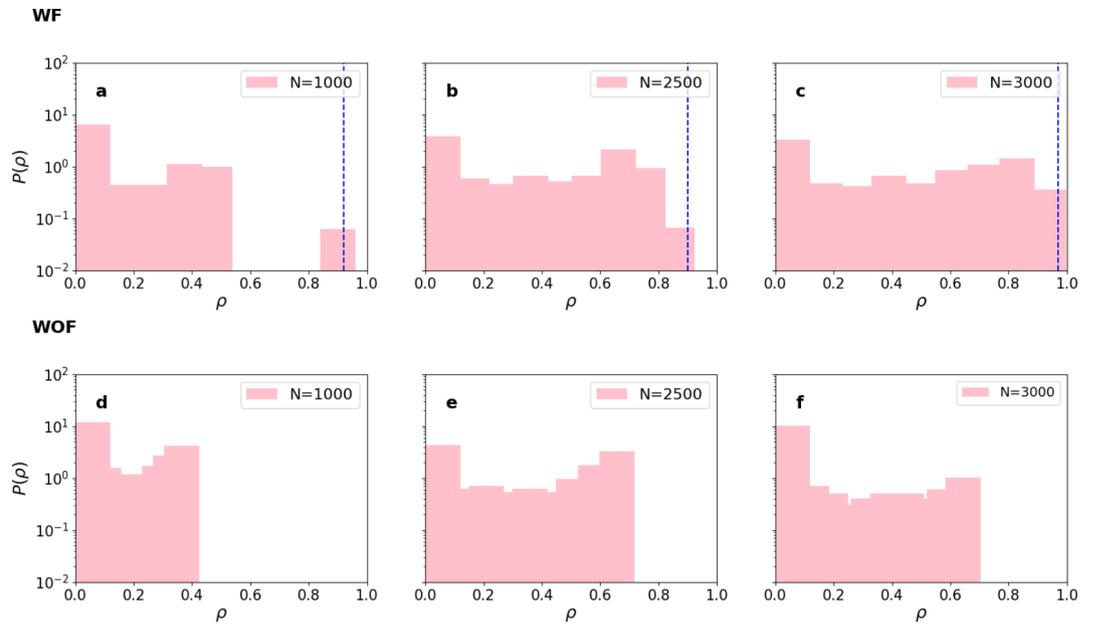

**Fig. 9**. The figures (**a**–**c**), (**d**–**f**) depict histograms of density at a specific time (t = 60τ), showing the distribution of individuals within different system sizes (N = 1000, 3000 and 4000) in the presence and absence of frontliners respectively. The blue dotted lines represent the density of the frontliners.

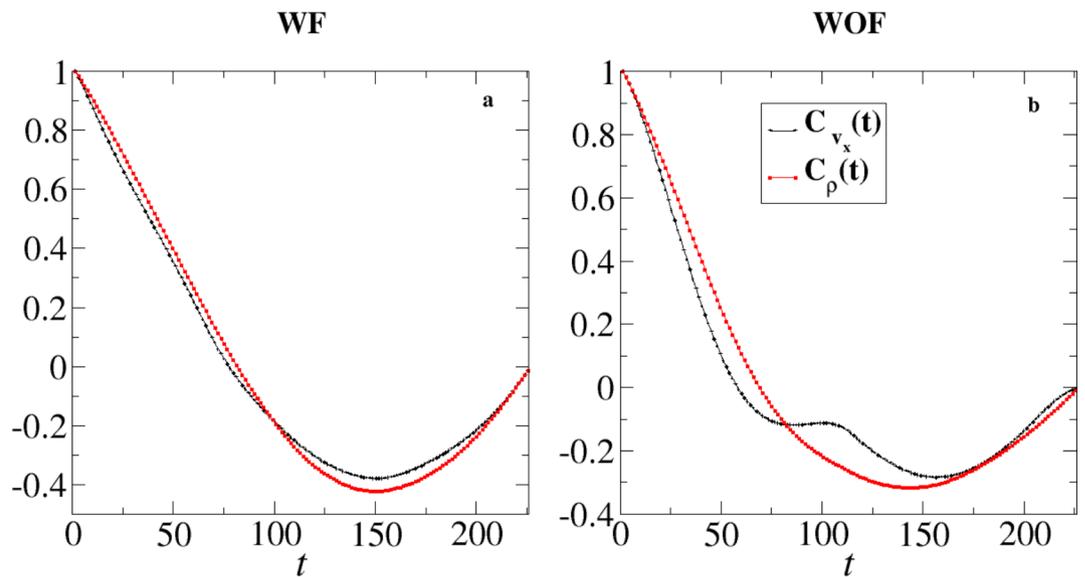

**Fig. 10**. The plots (**a**, **b**) depict the velocity auto-correlation ($C_{v_x}(t)$) vs. time (t) and density auto-correlation ($C_\rho$ (t)) vs. time (t) in the presence and absence of frontliners respectively.

The frontliners' density is marked with vertical dashed line in Fig. 9a–c. The shift in the mean density for both the cases is due to the different relaxation time for different N.

### Speed of density and velocity waves

To further quantify the characteristics of the traveling density and velocity wave, we calculate the auto-correlation functions of the fluctuations of density and velocity defined as; $C'_{v_x}(t) = \langle \delta v_x(t) \cdot \delta v_x(t + \delta t) \rangle$ and $C_\rho(t) = \langle \delta \rho(t) \delta \rho(t + \delta t) \rangle$ respectively. The fluctuations in velocity and density are defined as: $\delta v_x(t) = (v_x(t) - v')$ and $\delta \rho(t) = (\rho(t) - \rho_0)$, where $v'$ and $\rho_0$ is mean value of $v_x(t)$ and $\rho(t)$ over the time. The $C_{v_x}(t)$ and density $C_\rho(t)$ are averaged over 50 realizations and four different sizes of the race and many reference times $t_0$. Figure 10a,b show the plots of the auto-correlation functions of the density $C_\rho(t)$ and the velocity $C_{v_x}(t)$ for the system WF and WOF respectively. Both the $C_\rho(t)$ and $C_{v_x}(t)$ show the early time exponential decay with time. Both the curves almost overlap with each other. Hence, the velocity and density wave travels almost with the same speed. We also calculated the cross correlation of velocity and density at different times as shown in appendix A Fig. 12a,b for the system with and without frontliners. The nature of the correlation changes where the auto-correlation goes to negative values. Hence, the correlation between density and velocity weakens beyond that time and crowd becomes less cohesive.

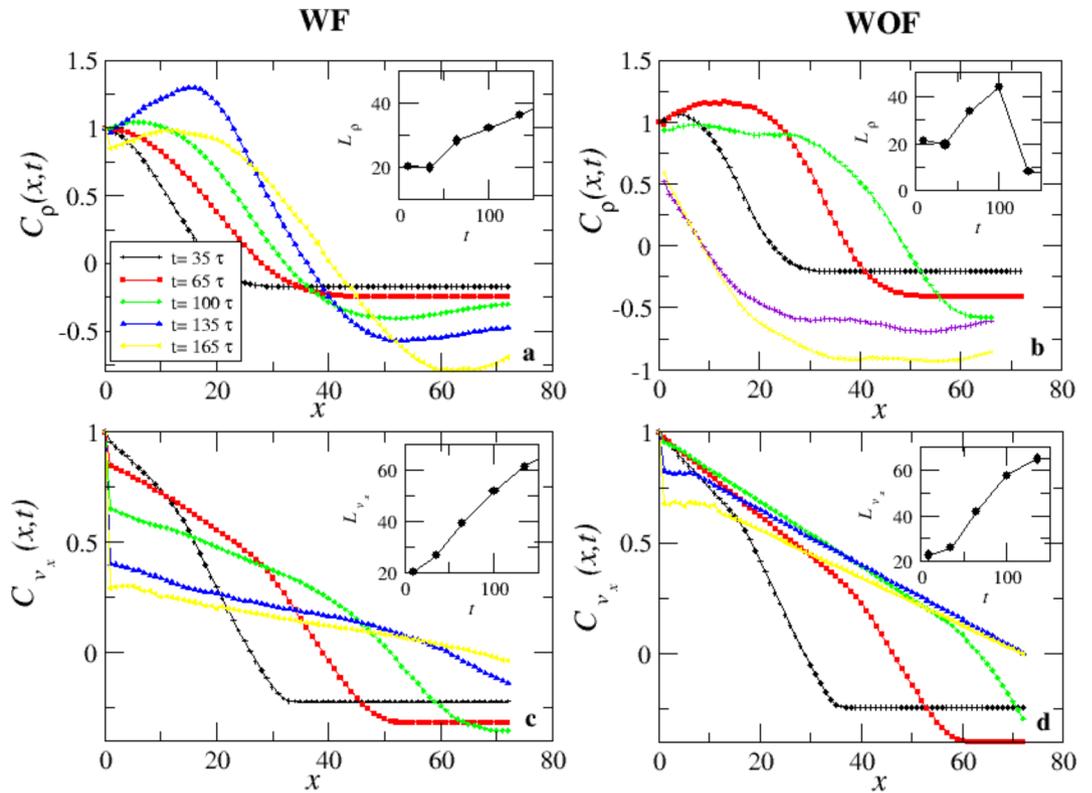

**Fig. 11**. The plots (**a**, **b**) showcase the spatial correlation functions for density $C_\rho(x,t)$ vs. $x$ for system ($N=1000$) WF and WOF in sequence. And plots (c-d) depict the spatial correlation functions for the velocity $C_{v_x}(x,t)$ vs. $x$ for the same. The insets of each plots show the corresponding characteristic lengths $L_\rho, v_x$ vs. time.

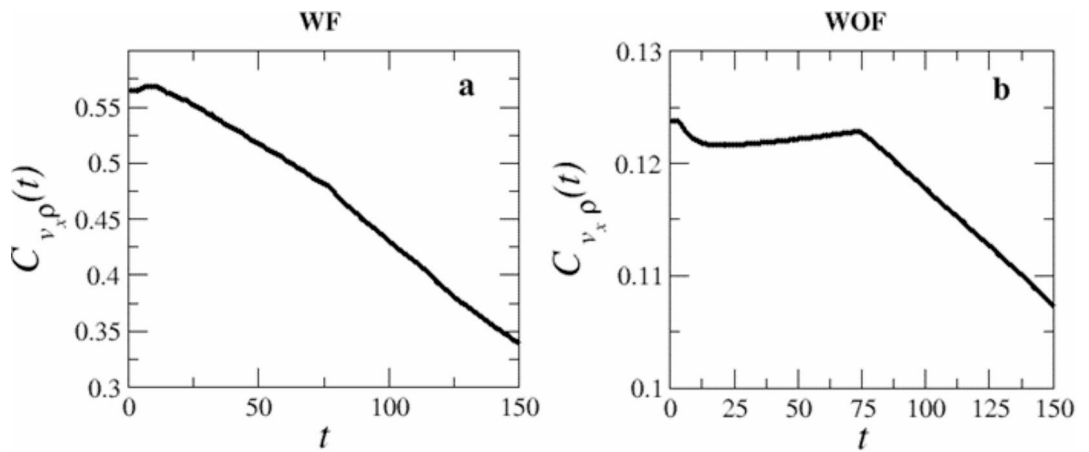

**Fig. 12**. The plots (**a**, **b**) depict the cross-correlation functions $C_{v_x\rho}(t)$ vs. time (t) in the presence and absence of frontliners respectively. Other parameters are the same as in Fig. 11.

### Discussion

We provide an agent-based model to replicate the crowd in a marathon race and examined the properties of the moving crowd. Through this study, we try to provide a supporting computational model to replicate the results obtained in recent experiment and hydrodynamic theory on real marathon races[34]. The races considered in[34], are guided by the race staff at the front of the race. Races without this oversight may be less organized but offer a more liberating experience. Hence, we introduce a model that incorporates both scenarios to accurately simulate the actual marathon experience. Most of the outcomes of our study, qualitatively fit with the observables which are reported in[34] for different races by using the empirical data obtained by video-tracking. Additionally our study introduce a model for race without any race staff and give the detailed comparison for the two cases. The main result of our paper is the propagation of hybrid coupled velocity-density wave in the direction opposite to direction of crowd motion as found in real races[34]. The speed distribution depicts the propagating wave has a constant speed irrespective of the number of participants in the races considered. That makes the characteristic of the travelling wave in polarised crowd very different from the usual wave moving through a medium, which is dispersive in nature. The distribution of longitudinal and transverse velocity shows that the fluctuation of velocity primarily occurs in the longitudinal direction and velocity fluctuations in transverse direction are highly suppressed, unlike the velocity fluctuation in polar flock[47–51]. We observed a non-monotonic relation between the local density and velocity in the system. This plot shows a largest density with a most-probable speed, which reflects the presence of density band in both the cases with and without frontliners. We also found that the key characteristics of the moving crowd are similar for both types of races. Only the contrast of density is weak for the system without frontliners due to the completely free motion in the direction of crowd propagation. The results reported in[34] mainly focused on the characteristics of velocity of participants. In our present study we also provide a detailed correspondence of density and velocity fields. Any information about the density of the participants and its relation with their velocity can be useful for the control and predict the behaviour of the system.

The appearance of the density wave in the system, is applicable for wide range of active systems starting from microscopic scales like collection of bacteria, molecular motors, etc. to the macroscopic scales like animal herd. The key result of our study: the presence of density wave is already reported in a system on a much smaller scale like collection of bacteria sliding on patterned substrate[41]. Our findings lead to future direction of research, focusing on examining the impacts of various types of boundaries positioned perpendicular to the passage. The present model we introduced here, is simplistic and ignored any kind of social or non-contact interactions among the participants. Inclusion of such forces can make the crowd propagation more robust. The detailed study of the system with non-contact forces can be an interesting future work. Another prospective pathway of future research involves the introduction of some external perturbations to investigate the response i.e. stampede like situation[18,52,53].

## Data availability
The datasets used and/or analysed during the current study available from the corresponding author on reasonable request.

## Appendix A: Appendix
Spatial Correlation Function and cross correlation function We have demonstrated the two-point spatial correlation of density and velocity $C_{\rho,v_x(x,t)}$ at different times as shown in Fig. 11a–d. The two-point spatial correlations are defined as $\langle C_\rho(x,t) = \delta \rho(x',t) \delta \rho(x+x',t) \rangle$ and $C_{v_x}(x,t) = \langle \delta v_x(x,t) \cdot \delta v_x(x+x',t) \rangle$, where the fluctuations $\delta \rho(x,t) = (\rho(x,t) - \rho_0(t))$ and $\delta v_x(x,t) = (v_x(x,t) - v'(t))$. The $\rho_0(t)$ and $v'(t)$ are the average density and velocity at time t. The average $<..>$ is performed over x', spanning quarter of the span of the crowd and 100 independent realisations of the system. The results are obtained for system size $N=1000$. We find that for the system WF, first with time the correlation increases and using the two-point correlation we measure a typical correlation length $L_{\rho,v_x}$, which are obtained by the first zero crossing of the correlation functions. For the system WF, the length monotonically increases for both the fields; density and velocity Fig. 11a,c (insets), whereas for the system WOF, it is non-monotonic with respect to time for the density field as shown in the inset of Fig. 11b, which is due to the less cohesion among the participants in the absence of frontliners in the system.

We further calculate the cross-correlation function defined as $C_{v_x\rho}(t) = <v_x(t) \rho(t)>$. It is an important quantity and it gives the correlation of density and velocity in the system. The cross correlation is defined by taking the product of local one-dimensional density $\rho(x,t)$ and velocity $v_x(x,t)$ and then averaging the product over whole span of the crowd. Hence, it gives a number at different times. The further averaging is performed over 50 independent realisations for the system of size $N=1000$. If both density and velocity are in the same phase, then the correlation will be high, else it will decay. The plot of cross-correlations for system WF and WOF is shown in Fig. 12a,b respectively. We find that for the WF, the correlation shows the very early time increment for $t \simeq (6-10) \tau$, this is the time when we start seeing clear density wave in the system, later correlation decay linearly with time. Further, the slope of linear decay changes near $t \simeq 75\tau$ which is due to the periodic nature of density and velocity waves in the system, at approximately the same time, the auto-correlation of density and velocity $C_{v_x\rho}(t)$ switches to negative values as shown in 10a. This is due to the change in the nature of density and velocity bands in the moving crowd. Further, we also showcase the plot of cross-correlation for the system WOF and found the same feature, first the correlation decreases, which is due to an initial spread of the crowd, and then it increases and reaches to maxima at $t \simeq 75\tau$. By this time, we observe a good correlation in density and velocity auto-correlation and then it switches to negative values as shown in 10b. Later, it decayed sharply with time due to gradual weakening of density and velocity correlation in the system.

The two-point correlations and cross-correlations we find for the moving participants is very different in nature as found in usual flocking system of self-propelling agents as given in[54].

### Acknowledgements

P.J. and S. M. thank Prof. Jacques Prost for useful discussions. P.J. gratefully acknowledge the DST INSPIRE fellowship for funding this project. The support and the resources provided by PARAM Shivay Facility under the National Supercomputing Mission, Government of India at the Indian Institute of Technology, Varanasi are gratefully acknowledged by all authors. S.M. thanks DST-SERB India, ECR/2017/000659, CRG/2021/006945 and MTR/2021/000438 for financial support. P.J. and S.M. also thank the Centre for Computing and Information Services at IIT (BHU), Varanasi.

### Author contributions
P.J. and S.M. conceptualize the problem. P.J and S.M. wrote the main manuscript text and P.J. prepared all figures and animations. All authors reviewed the manuscript. S.M. supervised the work.

### Declarations

### Competing interests
The authors declare no competing interests.

### Additional information
**Correspondence** and requests for materials should be addressed to P.J.

**Reprints and permissions information** is available at www.nature.com/reprints.

**Publisher's note** Springer Nature remains neutral with regard to jurisdictional claims in published maps and institutional affiliations.